# Computer Simulations of a Heterogeneous Membrane with Enhanced Sampling Techniques

Yevhen K. Cherniavskyi,[1] Arman Fathizadeh,[2] Ron Elber[2,3*], D. Peter Tieleman[1*]

[1] Department of Biological Sciences and Centre for Molecular Simulation, University of Calgary, 2500 University Drive NW, Calgary AB T2N 1N4, Canada

[2] The Oden Institute for Computational Engineering and Sciences, University of Texas at Austin, Austin, TX, 78712

[3] Department of Chemistry, University of Texas at Austin, Austin, TX, 78712

Corresponding authors: ron@oden.utexas.edu tieleman@ucalgary.ca

Abstract

Computational determination of the equilibrium state of heterogeneous phospholipid membranes is a significant challenge. We wish to explore the rich phase diagram of these multi-component systems. However, the diffusion and mixing times in membranes are long compared to typical times of computer simulations. To speed up the relaxation times, advanced simulation methods are used. We evaluate the combination of enhanced sampling techniques such as MDAS (Molecular Dynamics with Alchemical Steps) and MC-MD (Monte Carlo with Molecular Dynamics) with a coarse-grained model of membranes (Martini) to reduce the number of steps and force evaluations that are needed to reach equilibrium. We illustrate a significant gain compared to straightforward Molecular Dynamics of the Martini model by factors between three to ten. The combination is a useful tool to enhance the study of phase separation and the formation of domains in biological membranes.

1. Introduction

Biological membranes are complex constructs that function as a semi-permeable barrier between the cell interior and the external environment. It consists of phospholipids, cholesterol and protein molecules, and more. The membrane components assemble into microphases and nanodomains and regulate cell function. According to the raft hypothesis, lateral inhomogeneity of the lipid membranes plays a key role in cell signaling, protein aggregation, and membrane fusion.[1]

A number of experimental techniques such as X-ray and neutron scattering [2, 3], NMR [4], and others [5] provide structural data on lipid membranes. Despite significant progress, studying the structure of biological membranes at molecular resolution remains a challenging task due to the disordered and fluid characteristics of these systems. Scattering-based techniques can directly probe the spatial organization of lipid membranes without introducing additional probes that alter the membrane structure [6]. However, the scattering signal is a spatial and temporal average over many fluid structures, leading to a smooth and less detailed signal. The averaging and separation of signals is even more difficult in mixed membranes with multiple types of phospholipids.[7]

The rapid growth in the power of computers and the development of simulation methodology increase their use for the study of lipid membranes [8-10]. Recently, a state-of-the-art realistic model of the plasma membrane that contains more than 60 types of phospholipids was developed. [11] Nevertheless, sampling the equilibrium distribution of constituents of heterogeneous membranes remains computationally hard. The prime problem is the slow diffusion of phospholipids in the membrane plane that prevents efficient mixing at time scales accessible for Molecular Dynamics (MD). One approach to reducing the computational cost is to use coarse-grained description, such as the Martini model.[12] On average, the Martini model represents four heavy atoms as a single particle or a bead. The reduction in the total number of particles compared to atomistic force fields leads to a significant gain in computational resources. Moreover, the removal of the fast degrees of freedom (e.g., vibrations of atomic bonds), enables the use of larger time steps and diminishes thermal noise and friction. The diffusion coefficient is $D=k_B T/\gamma$ where $k_B$ is the Boltzmann constant, T is the temperature, and $\gamma$ the friction coefficient. Hence the diffusion is faster when the friction coefficient is smaller. The effective speedup of the diffusion within the Martini model (~4-5 times faster than in atomistic models and experiments) was documented in the literature.[12-14] However, even with the significant speedup compared to atomistic models, reaching the equilibrium of large heterogeneous membranes with classic Molecular Dynamics (MD) and the Martini force field is computationally expensive.

Recently a sampling approach that combines Molecular Dynamics (MD) with Monte Carlo (MC) approaches in the grand canonical ensemble was proposed. The method is particularly suitable for the simulation of heterogeneous lipid membranes in which the different lipids are quite similar.[15] The system is sampled by alternating steps of (1) a straightforward MD step in the microcanonical or canonical ensemble and (2) an MC move that replaces a phospholipid by another phospholipid of a different type. A random lipid is selected for an adjustment and if the MC move is accepted following the usual Metropolis criterion, the lipid molecule is modified to its lipid counterpart (for example, from DPPC to DPPS). The MD/MC approach is not bound by the slow lateral diffusion of lipids, and phase separation and mixing are sampled more efficiently than straightforward MD. The challenge with the MC approach is that the lipid types must be similar for the MC move to be accepted with a reasonable probability. Because the acceptance is typically low in MC-MD in membranes, many trials are needed. To increase the number of trials for a fixed number of force evaluations only a single or a few MD steps separate two MC steps. The extraction of kinetic information (such as the diffusion constant) is not possible in this approach. The rapid flips between MD and MC moves may also lead to hysteresis, and the sampling may deviate from the desired equilibrium.

Generating trial MC moves with higher acceptance probabilities allows longer MD trajectories between the MC moves and better relaxation to equilibrium. It also makes it possible to extract short time kinetic information, such as local diffusion constant. Therefore, a new algorithm was proposed: Molecular Dynamics with Alchemical Steps (MDAS).[16] Instead of performing the exchange in a single MC move, we conduct a gradual growth of the two lipids into their counterparts, relying on the Jarzynski equality [17] and the algorithm for candidate Monte Carlo move [18] to obtain the correct statistics. This exact approach significantly increases the acceptance probability of the MC move. Instead of modifying a single phospholipid, an exchange of a pair of phospholipids of different types is considered. The exchange of two molecules instead of just one keeps the composition of the phospholipids fixed.

The MDAS algorithm generates steps that are more likely to be accepted than conventional MC moves. However, if the interactions of the exchanged phospholipids with their environments are significantly different, the gradual modification of the molecules can be inefficient. An example of a challenge for atomically detailed simulation, which is discussed in **Results**, is the pair of PS and PC phospholipids. PC is neutral, while PS is negatively charged. Therefore, the electrostatic interactions impact the rate of relaxation to equilibrium, leading to many rejected MC-MD and MDAS steps. This problem is not present in the coarse-grained model of Martini. The mutation of PS to PC in Martini adjust one bead (the head group) with only short-range interactions (Figure 1). The challenges with atomically detailed simulations, even with the enhanced sampling technique of MDAS, makes the use of a coarse-grained model attractive for the PS and PC system.

Here we explore the use of MDAS and MC-MD algorithms for the sampling of a lipid mixture with the Martini force field. We test the performance of the algorithm with a binary DPPC/DPPS mixture and benchmark it against straightforward MD and MC-MD approaches.

2. Theory

**MDAS and MC-MD**

In the current paper, we use two approaches for the simulation of the system – MC-MD and MDAS. The MC-MD simulation is conducted as a series of alternating straightforward MD steps and MC lipid exchange moves. We randomly select a pair of different types of phospholipid molecules (e.g., one DPPC and one DPPS molecule) and change the chemical identity of each of the headgroups (bead) to the other type. We do not modify coordinates (Figure 1), only the corresponding potential parameters. To swap DPPC with DPPS we change the headgroup type P5 of DPPC to bead type Q0 of DPPS. Then, we either accept or reject the proposed MC move based on Metropolis criterion:

$$P_{accept} = min\left[1, exp\left(\frac{-\Delta U}{kT}\right)\right] \qquad (1)$$

$$\Delta U = U(r_1, \ldots, r_i, \ldots, r_j, \ldots) - U(r_1, \ldots, r_j, \ldots, r_i, \ldots)$$

where ΔU is the energy difference before and after the trial move, $r_i$ and $r_j$ are the coordinates of the selected pair of lipids, k is the Boltzmann constant, and T is the temperature. If the move is accepted, we continue with an MD trajectory starting from the new state. If the move is rejected, we go back to the state before the exchange attempt and run a straightforward MD step.

The MDAS algorithm substitutes a single-step MC move with a gradual adjustment, which is computed using an alchemical trajectory (AT).[16] During the AT, the selected phospholipid molecules are changing into their counterparts (e.g., DPPC to DPPS and vice versa). MDAS simulation consists of (1) straightforward MD trajectories for sampling, and (2) "alchemical" trajectories (AT) that modify the phospholipids and generate candidate MC moves.[18] The AT is similar to alchemical methods used to determine the free energy difference between two states.[19, 20] Like in a free energy calculation, the AT path is parameterized with λ∈[0,1]. When λ=0 we are at the beginning of the attempted exchange, and when λ=1, at the end of it. Correspondingly the potential energy along the AT is $U(R, \lambda)$. The dependence of the potential on λ is the choice of the user. The simplest implementation is linear. For an exchange of a system A to a system B we have $U(\lambda) = (1 - \lambda)U_A + \lambda U_B$. We conduct the AT as follows: We run M steps of straightforward MD at a fixed value of λ, and then increase λ by a small Δλ in a single step. Starting with λ=0 the M steps and the increase in λ are repeated until λ is equal 1. The work done on the system during the entire AT is:

$$w(\lambda = 0 \to \lambda = 1) = \sum_i \left( U_{\lambda_i + \Delta\lambda}(x_i) - U_{\lambda_i}(x_i) \right) \quad (2)$$

Where $x_i$ are the system coordinates after $i$ repeats of the M steps. The total work is used in an acceptance-rejection criterion of the AT move [16] similar to Metropolis (Eq. (1)):

$$P_{accept} = min\left[1, exp\left(\frac{-w}{kT}\right)\right] \quad (3)$$

If the move is accepted, we continue the simulation from the last configuration of the state λ=1 (the lipids are exchanged) and proceed with another segment of a straightforward MD trajectory. If the move is rejected, we discard the AT and continue from the last step of the previous straightforward MD segment.

Thus, the entire MDAS simulation consists of a series of short conventional MD trajectories and exchange steps (AT) in between. We use only the conventional MD segments to calculate the thermodynamic properties of the system.

A specific MDAS choice of Δλ=1 and M=0 is equivalent to a single-step MC exchange. In MDAS, the role of the AT trajectory is to allow the system to relax. A gradual exchange with Δλ ~0.001-0.01 yields work significantly lower compared to the direct MC exchange, as it produces less steric overlaps and bad contacts. Therefore, the MDAS steps in atomically detailed models are accepted with a much higher probability than direct MC. However, if the two types of phospholipids are similar (as DPPC and DPPS are in Martini), the additional cost of computing AT versus direct MC is not necessarily beneficial.

3. **Methods**

The simulations were performed with the standard Martini 2.2 force field [21] with and without polarizable water. [22] Electrostatic interactions were modeled with the reaction-field method. [23] The screening constant was 15 and 2.5 for non-polarizable and polarizable water, respectively. The cutoff distance of the vdW interactions was at 1.1 nm with a potential-shift modifier. All the simulations were conducted at 335 K. The temperature was fixed with velocity rescaling [24] with 1.0 ps coupling constant and two separate coupling groups for the membrane and the solvent. A semi-isotropic (xy and z directions) Parrinello-Rahman barostat [25] maintains a constant pressure of 1.0 bar with the coupling constant of 12.0 ps. Standard MD simulations were performed with GROMACS 2019.1 [26, 27]

MDAS and MC-MD simulations were conducted with an in-house modification of GROMACS 2019.1. A 1:1 mixture of DPPC and DPPS was considered. Because the differences in Martini between DPPC and DPPS are small (Figure 1), a short AT of 1000 steps was sufficient. The parameter λ is modified every ten steps, hence, Δλ=0.01. After 1000 steps, the total work is computed and the proposed move is accepted or rejected. Then we sample another 2000 steps of straightforward MD before attempting another AT. The same approach was used for MC-MD sampling scheme, but the AT is a single step (Δλ=1.0). To compare different methods on equal footing, we consider the number of force evaluations used per number of sampled configurations. The cost of a single attempt of MDAS exchange is 3000 force evaluations (2000 straightforward MD and 1000 AT steps), and it is 2001 force evaluations for a single exchange attempt in MC-MD.

For comparison, we also run an atomistic MDAS simulation for the DPPS/DPPC system using the program NAMD [28] and the CHARMM36 force field. [29] We had considerable success in the past, in simulating the mixture of DOPC and DPPC.[16] We show in Figure 2 the required alchemical changes in the atomically detailed MDAS simulations. There are 26 atoms that require modifications, which is a considerably more complex task than the single-particle exchange of Martini. As noted earlier, changes in electrostatic within the atomic models pose an additional and significant challenge to MDAS. The assigned charges of the phospholipids are of the CHARMM 36 force field [30] that was used in the atomistic simulations. We consider a 1:1 binary mixture of DPPC and DPPS. The bilayer consists of 200 phospholipids and is solvated with TIP3P water molecules.[31] One hundred potassium ions were added to neutralize the system. The entire system contains approximately 50,000 atoms. The membrane was first equilibrated in the NPT ensemble for 10ns, which was followed by 10ns NVT simulation. To examine the efficiency of MDAS, we conducted 100 AT attempts. Each AT was for a total length of 100 ps with $\Delta\lambda = 0.001$. To avoid the so-called end-point catastrophe, we used a soft-core potential with the following form to treat Van Der Waals interactions during the exchange [32]:

$$U_{soft}(r_{ij}) = 4\varepsilon\lambda\left[\left(\frac{\sigma_{ij}}{r_{ij}^2 + \delta(1-\lambda)}\right)^6 - \left(\frac{\sigma_{ij}}{r_{ij}^2 + \delta(1-\lambda)}\right)^3\right] \qquad (4)$$

Where δ=5.0 and λ changes from 0 to 1 during the alchemical step. Note that at λ=1 the above expression turns into 6-12 Lennard-Jones and the interaction vanishes at λ=0. The timestep was 1fs in all the atomistic simulations.

## 4. Results

We use the 1:1 DPPC/DPPS lipid mixture as our test system for sampling efficiency. We simulate the Martini model using MDAS, mixed MD-MC, and straightforward MD. The atomically detailed calculations were attempted with MDAS. The initial state of the system was of separated DPPC and

DPPS molecules (Figure 3), which is far from the equilibrium of a uniformly mixed membrane. The radial distribution function g(r) of the PO4 beads of DPPS (or DPPC) phospholipids monitors mixing as a function of time. We have shown in [16] that the highest peak of the pair correlation function max[g(R)] is a good measure of the relaxation and is comparable to the alternative measure of mixing entropy.[33] As the mixing occurs, max[g(r)] approaches a constant value of uniform mixing of the two phospholipid types. To obtain a quantitative estimate of the relaxation rate, the evolution of max[g(r)] is fitted with an exponential function.

With the current choice of parameters for MDAS moves, the acceptance probability is about 29%. Figure 4 shows the time evolution of max[g(r)] for DPPS-DPPS PO4 beads as a function of the number of force evaluations. The fit of an exponential function to the evolution of max[g(r)] gives ~11.3 times speedup for the system mixing compared to straightforward MD. If we use the mixed MD-MC sampling scheme, which is equivalent to MDAS with an AT of a single step, the acceptance probability is about 16 %. An exponential fit of max[g(r)] as a function of the number of force evaluations suggests ~10.1 speedup of the mixing dynamics compared to sampling by straightforward MD. Figure 5 shows snapshots of the top view of the lipid bilayer simulated with MD and MDAS. After two million force evaluations in a straightforward MD simulation, the bilayer is far from laterally homogenous. At the same time, after 660 MDAS steps, which correspond to 2 million force evaluations, the two lipid types are well mixed.

To further explore the performance of the different sampling schemes with different parameterization, we simulate the same Martini system with polarizable water. As in the previous case, we benchmark MDAS and MD-MC methods against straightforward MD. With MDAS we obtain 9% acceptance probability for the DPPC/DPPS exchange move, which translates into ~2.5 times speedup compared to straightforward MD (Figure 6). After 10,000 attempts, we did not accept a single exchange move with MD-MC sampling. From the work values we estimate the average acceptance probability as $8.8 \times 10^{-5}$.

Finally, we also attempted to simulate the mixing of the DPPC/DPPS system using an atomically detailed MDAS model. We evaluate 100 AT steps of length of 100 picoseconds (each) using $\Delta\lambda =$

0.001. The length of the straightforward MD trajectories between AT attempts was 100 ps as well. In figure 7, we show a histogram plot of the work values evaluated from the AT trajectories. The distribution is broad and includes high work values, which makes the acceptance probability less than $10^{-5}$ and impractical for the current AT path.

## 5. Discussion

The Martini model offers an efficient approach to sample membrane configurations by reducing the number of particles and using softer energy landscapes compared to atomistic models. It enables the study of heterogeneous membranes, assembly, and separation. However, the enormous diversity of biological membranes and their shear sizes pose a significant challenge for converging straightforward MD simulations, even with the Martini model. It is for that reason that the combination of Martini and MDAS is promising. MDAS was illustrated to be a useful algorithm for atomically detailed simulations. It provides speedup of ~1,000 for specific systems.[16] However, there are lipid compositions that are difficult to simulate with atomically detailed MDAS models. The challenge in MDAS simulations is the design of the AT, such that the amount of work will be minimal and lead to significant acceptance probability. For example, an efficient acceptance probability when using ~100 picosecond trajectories of straightforward Molecular Dynamics is about 10%. This design is difficult for the exchange of phospholipids with different charges, as we illustrated in this manuscript for the DPPC/DPPS system. The charge and the membrane electric field in atomically detailed models relax slowly to the new equilibrium imposed by the exchange. Lack of significant electrostatic interactions is another advantage of Martini compared to atomistic models. The exchange pathways are simple to design and the overall variations in the energy landscape are smaller in Martini compared to atomically detailed models. Therefore, more AT are accepted. The design of an efficient AT for diverse pairs of phospholipids is a topic of ongoing research.

We illustrate these observations numerically using straightforward MD and MDAS calculations for an atomistic and the Martini models. The result of the evaluation is that MDAS has the potential to be significantly advantageous to straightforward MD. The efficiency of the MDAS algorithm can be evaluated by exploratory short ATs. The distribution of the computed *w-s* in these exploratory runs assesses if MDAS is efficient or not. Acceptance probability of order of or greater than 10% allows for straightforward MD trajectories of about 100 picoseconds long between ATs. If the acceptance probability is below the threshold, we can always return to straightforward MD.

What types of AT generate small values of work? In our experience, a modification of the hydrocarbon chain (lengths, or single and double bond along the lipid chain) are good candidates for an MDAS calculation in atomic details. A modification of the head group is more challenging for atomically detailed models compared to Martini. For phosphate head groups of different charges, an efficient AT is hard to find. Similarly, an inefficient MC-MD move is found in the Martini model that incorporates electrostatic of water.

The DPPC/DPPS Martini system with polarizable water is an interesting example in which the MC-MD algorithm (or a single step AT) is not efficient. In the non-polarizable case, the charged PS headgroups interact electrostatically only with the ions as the water beads in the Martini model are not charged. However, more electrostatic interactions are present when the polarizable water model is used. The water model includes an induced dipole that interacts with the charges of the head group. When we switch from PC to PS in a single step, the electrostatic interactions of the PS groups with water molecules contribute to a significant energy difference between the exchanged states. As a result, the average acceptance probability in MC-MD is $\sim 8.8 \times 10^{-5}$. However, if we simulate the transition between PS/PC headgroup gradually with an AT, the polarized water molecules have sufficient time to lose their transient dipoles, and the work of the transition is reduced, which translates into higher acceptance probability of the proposed exchange move.

An interesting application of sampling methods of phospholipid mixtures is an investigation of asymmetric lipid bilayers. One can propose an AT that would exchange lipids between different layers.

Since flip/lop movements between bilayers is an activated process, such a move can greatly increase the equilibration rate. Of course, one should keep in mind that membrane asymmetry is maintained in biological systems, and a true equilibrium may not be desired in such cases. However, for the investigation of synthetic systems, which may still be asymmetric, the MDAS algorithm and the Martini model are promising.

**Conclusions**

In the current paper, we explored the possibility of using exchange-based MDAS and MC-MD algorithms for the efficient sampling of mixed lipid bilayers within the Martini model. The model system is a binary 1:1 DPPC/DPPS mixture that illustrates how advanced sampling approaches within the Martini force field can significantly increase the sampling and open new possibilities for simulations of large multi-component lipid mixtures. For the relatively small system considered in this paper (400 lipid molecules) the speedup factor can be as large as eleven. It is concluded that for exchange moves with small energy modifications, the use of MC-MD sampling scheme might be a desirable approach. However, for more complex exchange moves that require substantial adjustments of the system, MDAS algorithm is superior to MC-MD approach.

Acknowledgements

Works in DPTs group is supported by the Natural Sciences and Engineering Research Council (Canada) and the Canada Research Chairs Program. Calculations were carried out on Compute Canada resources, funded by the Canada Foundation for Innovation and partners. Membrane research in RE laboratory is supported by an NIH grant GM 111364, an NSF grant BIO-1815354, and a Welch grant F-1896.

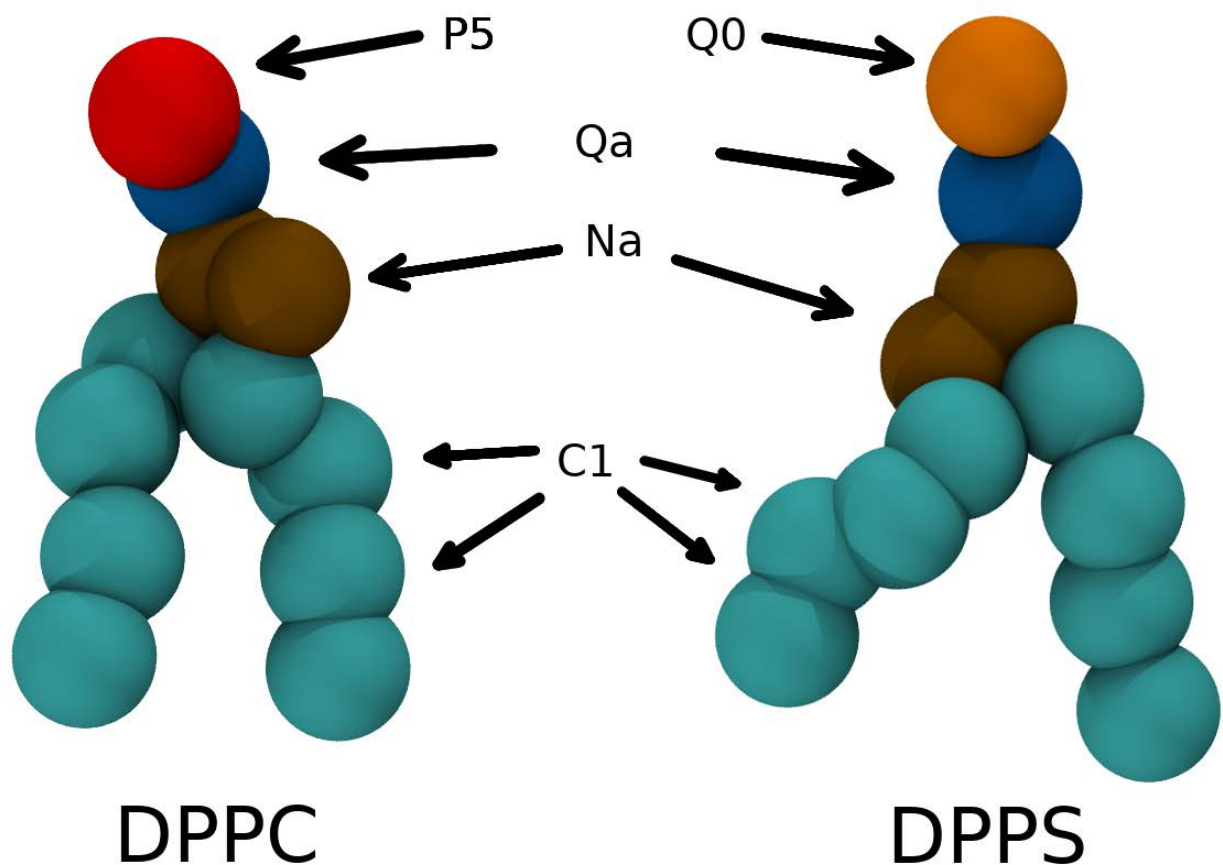

Figure 1. DPPC and DPPS lipid molecules in the Martini model. On average 4 heavy atoms with corresponding hydrogens are represented as one bead. Different bead types are marked on the figure and highlighted with different colors. The only difference between DPPC and DPPS lipid molecules in the Martini force field is the type of one headgroup bead.

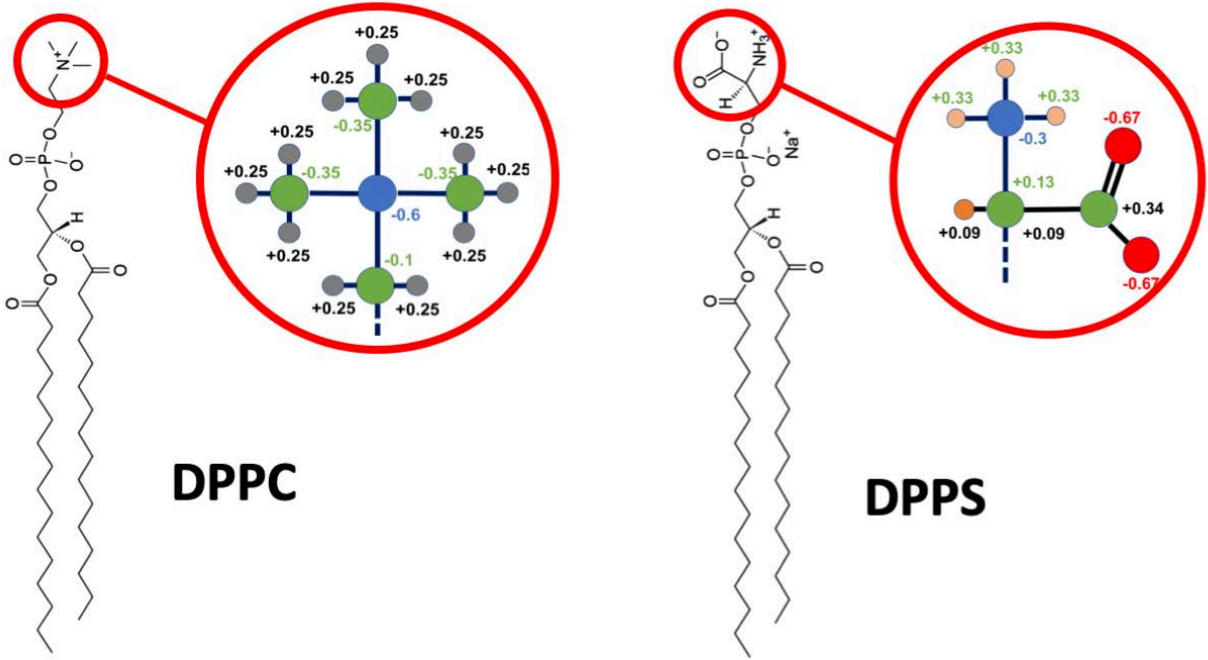

Figure 2. Structures of DPPC and DPPS lipid molecules with required alchemical changes in the atomically detailed MDAS simulations.

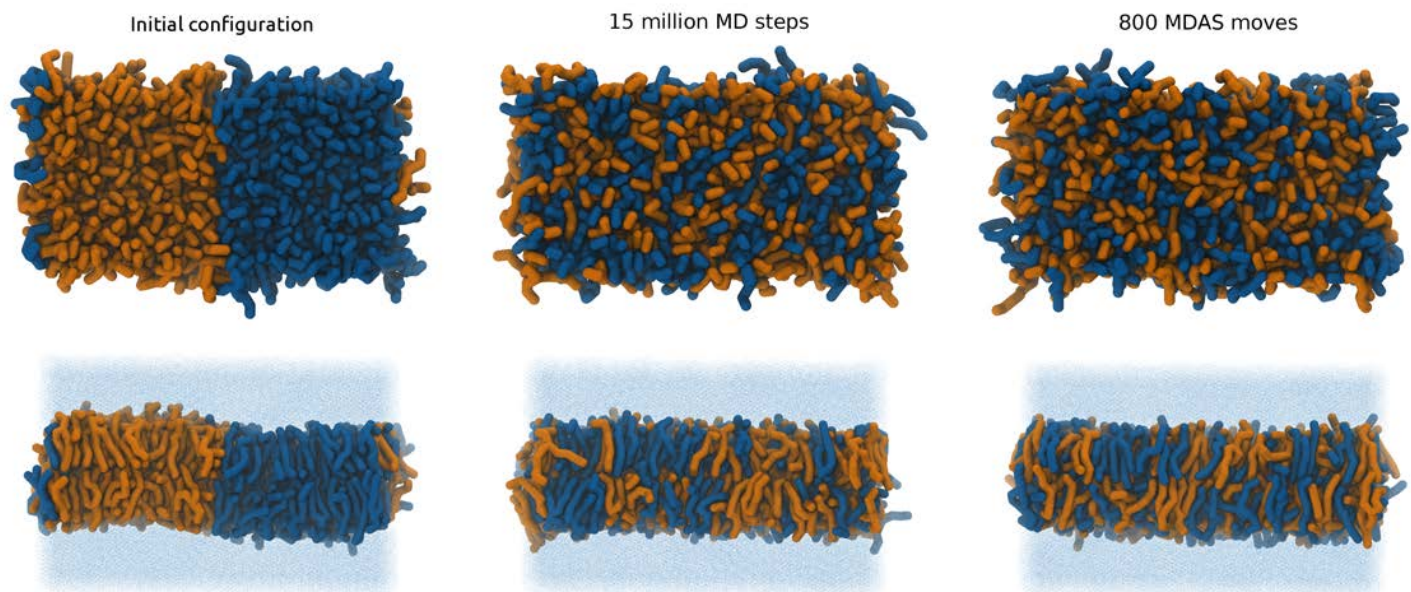

Figure 3. Top and side view of the DPPC/DPPS bilayer used as a test system. DPPC lipid molecules are shown in dark blue and DPPS lipid molecules in orange. Water is represented as blue dots.

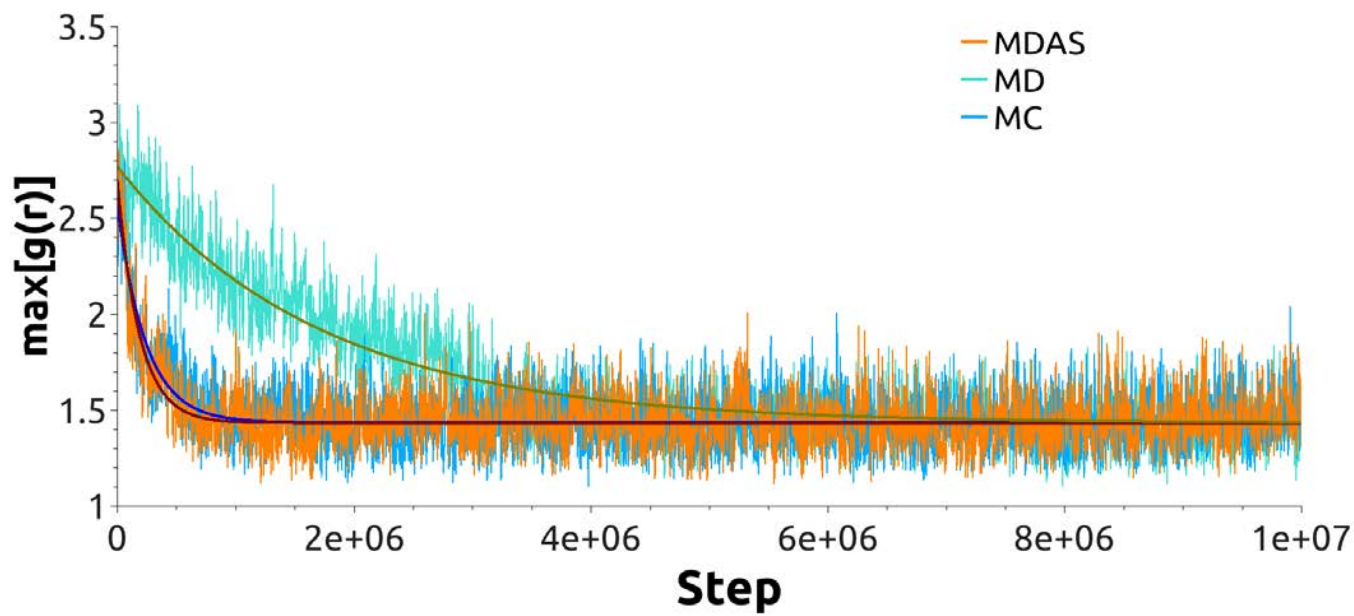

Figure 4. The evolution of max[g(r)] for the DPPS PO4 beads for the system with non-polarizable Martini water. Orange line – system sampled with MDAS algorithm; turquoise – straightforward MD; light blue – MC-MD approach. Solid dark green (MD), wine (MDAS) and blue (MC-MD) lines represent exponential fits.

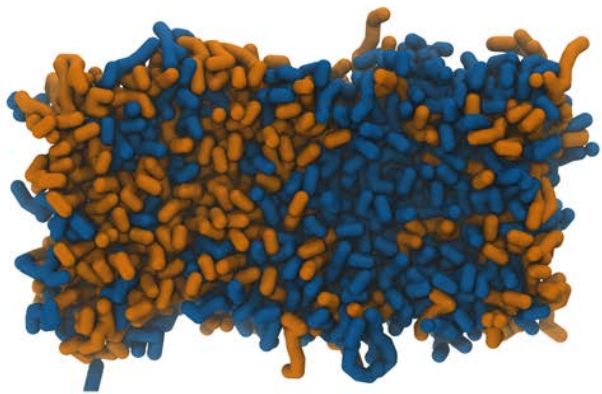 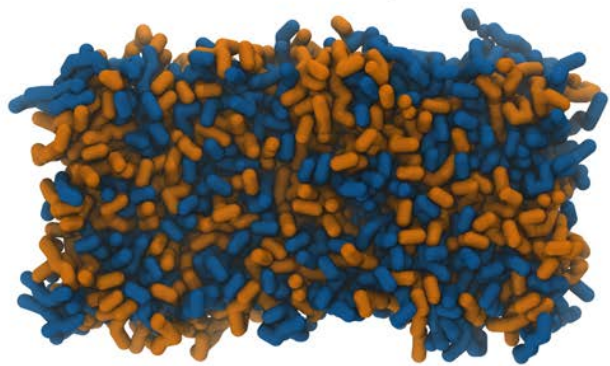

Figure 5. Top view of the DPPC/DPPS bilayer after 2 millions MD steps and 660 attempted MDAS moves. DPPC lipid molecules are shown in dark blue and DPPS lipid molecules in orange. Water molecules are not shown for clarity.

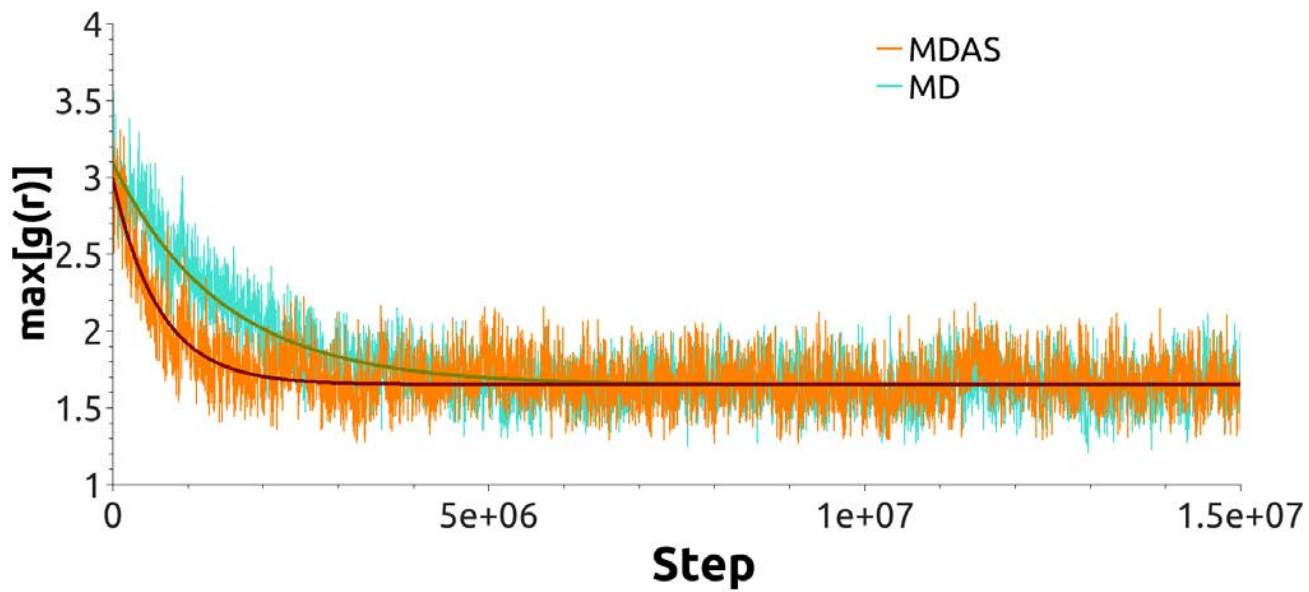

Figure 6. The evolution of max[g(r)] for the DPPS PO4 beads for the system with polarizable Martini water. Orange line – system sampled with MDAS algorithm; turquoise – straightforward MD. Solid dark green (MD) and wine (MDAS) lines represent exponential fits.

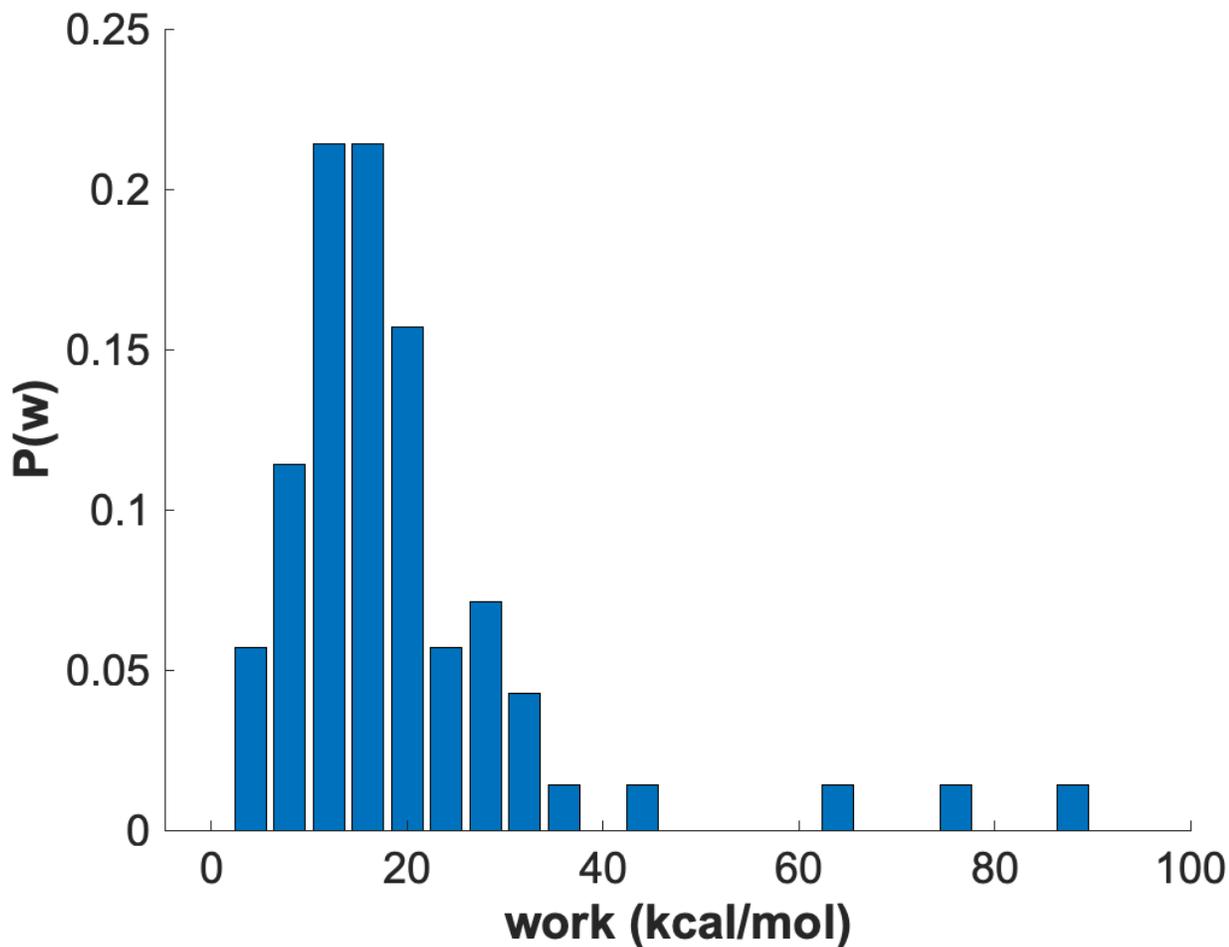
Figure 7. Histogram of the work values evaluated from alchemical trajectories.